\documentclass[conference,compsoc]{IEEEtran}

\usepackage{cite}
\usepackage{url}            
\usepackage{booktabs}       
\usepackage{amsfonts}       
\usepackage{nicefrac}       
\usepackage{microtype}      
\usepackage{lipsum}		    
\usepackage{doi}
\usepackage{subcaption}
\usepackage{pifont}
\usepackage{colortbl}
\usepackage{float}
\usepackage{balance}
\usepackage{xspace}
\usepackage{acronym}
\usepackage{xcolor}
\usepackage{adjustbox}
\usepackage{amssymb}
\usepackage{siunitx}
\usepackage{tabularx}
\usepackage{enumitem}
\usepackage{multirow}
\usepackage[subtle]{savetrees}
\usepackage{hyperref}   
\newcommand{\wcircle}[1]{\ding{\numexpr171 + #1}}
\newcommand{\bcircle}[1]{\ding{\numexpr181 + #1}}

\usepackage[strict]{changepage}
\usepackage{framed}

\definecolor{lightblue}{HTML}{B7E0FF} 
\definecolor{deepblue}{HTML}{024CAA}

\definecolor{formalshade}{rgb}{0.85,1,0.85} % Light green background
\definecolor{darkblue}{rgb}{0.0,0.6,0.30}   % Dark green border

\newenvironment{answer}{%
  \MakeFramed{\advance\hsize-\width\FrameRestore}%
  \noindent\hspace{-4.55pt}% Disable indenting the first paragraph
  \begin{adjustwidth}{}{7pt}%
}
{%
  \end{adjustwidth}\endMakeFramed%
}

\begin{document}
\title{Evaluating Large Language Models in detecting Secrets in Android Apps}

\author{
    \IEEEauthorblockN{Marco Alecci, Jordan Samhi, Tegawendé F. Bissyandé, and Jacques Klein}
    \IEEEauthorblockA{
        SnT, University of Luxembourg, Luxembourg \\
        Email: \{marco.alecci, jordan.samhi, tegawende.bissyande, jacques.klein\}@uni.lu
    }
}

\maketitle

\begin{abstract}

Mobile apps often embed authentication secrets, such as API keys, tokens, and client IDs, to integrate with cloud services. However, developers often hardcode these credentials into Android apps, exposing them to extraction through reverse engineering. Once compromised, adversaries can exploit secrets to access sensitive data, manipulate resources, or abuse APIs, resulting in significant security and financial risks. Existing detection approaches, such as regex-based analysis, static analysis, and machine learning, are effective for identifying known patterns but are fundamentally limited: they require prior knowledge of credential structures, API signatures, or training data. 

In this paper, we propose SecretLoc, an LLM-based approach for detecting hardcoded secrets in Android apps. 
SecretLoc goes beyond pattern matching; it leverages contextual and structural cues to identify secrets without relying on predefined patterns or labeled training sets. Using a benchmark dataset from the literature that covers over \num{5000} Android apps, we demonstrate that SecretLoc detects secrets missed by regex-, static-, and ML-based methods, including previously unseen types of secrets. 
In total, we discovered 4828 secrets that were undetected by existing approaches, discovering more than 10 "new" types of secrets, such as OpenAI API keys, GitHub Personal Access Tokens, RSA private keys, and JWT tokens, and more.

We further extend our analysis to newly crawled apps from Google Play, where we uncovered and responsibly disclosed additional hardcoded secrets. Across a set of 5000 apps, we detected secrets in 2124 apps (42.5\%), several of which were confirmed and remediated by developers after we contacted them.

Our results reveal a dual-use risk: if analysts can uncover these secrets with LLMs, so can attackers. This underscores the urgent need for proactive secret management and stronger mitigation practices across the mobile ecosystem.
\end{abstract}
\pagestyle{plain}
\section{Introduction}
\label{sec:introduction}

Mobile applications (apps) increasingly depend on cloud services for features such as data storage, authentication, and many others, making their integration nearly ubiquitous. To enable such integrations, developers frequently rely on authentication secrets (e.g., API keys, access tokens, client IDs) embedded within their apps. Nevertheless, a recurring problem is that developers often hardcode these secrets directly into Android apps, leaving them vulnerable to extraction through reverse engineering, as confirmed by existing studies~\cite{li2024automaticallydetectingcheckedinsecrets,zuo2019does,wei2025far}. Once exposed, adversaries can exploit these secrets to access sensitive data, manipulate or corrupt resources, incur financial costs through unintended API usage, or cause denial of service (DoS) by exhausting invocation limits~\cite{li2024automaticallydetectingcheckedinsecrets}, therefore representing a known vulnerability: “CWE-798: Use of hard-coded credentials"~\cite{CWE798}.

While most existing tools focus broadly on code and code-sharing platforms, Li et al.~\cite{li2024automaticallydetectingcheckedinsecrets} recently conducted a systematic categorization of secret-detection approaches applicable to Android apps, covering more than ten existing tools~\cite{meli2019bad, gitleaks, repo-supervisor, trufflehog, whispers, ggshield, git-secrets-hunter, zuo2019does, feng2022automated, wen2022secrethunter, han2023credential, saha2020secrets, lounici2021optimizing}.
The approaches were classified into three main families of techniques: \wcircle{1} \emph{intrinsic-value-based analysis}, which relies on properties such as entropy and regex matching; \wcircle{2} \emph{static analysis}, which traces data flows through APIs; and \wcircle{3} \emph{machine-learning-based methods}, which classify candidate strings based on contextual features.
They then tested one approach from each category (specifically, Meli et al.~\cite{meli2019bad} for intrinsic-value analysis, \texttt{LeakScope}~\cite{zuo2019does} for static analysis, and \texttt{PassFinder}~\cite{feng2022automated} for machine-learning-based analysis) on a dataset of \num{5135} Android apps, discovering \num{2142} secrets affecting \num{2115} different apps~\cite{li2024automaticallydetectingcheckedinsecrets}.
While each family of detection techniques has its own advantages, they all share a critical limitation: \textbf{they require knowing exactly what to look for in advance}. Regex-based detectors~\cite{meli2019bad} rely on carefully crafted patterns (e.g., Google API keys with fixed structure \texttt{AIza[0-9A-Za-z\-\_]{35}}) static analysis tools~\cite{zuo2019does} depend on predefined API signatures, and machine-learning methods~\cite{feng2022automated} require training data that may not capture novel or obfuscated secrets. As a result, all three approaches can fail to detect previously unknown or cleverly hidden authentication credentials.

This observation raises a question: \emph{To what extent can large language models (LLMs) help overcome this limitation?} LLMs are trained to capture both the intrinsic structure of strings and their surrounding semantic context, which may allow them to infer when a value is a secret even without prior knowledge of what to look for. In this paper, we explore this possibility by proposing \texttt{SecretLoc}: a novel LLM-based approach for detecting hardcoded secrets in Android apps.

On the same dataset of apps analyzed by Li et al.~\cite{li2024automaticallydetectingcheckedinsecrets}, our approach detected secrets that were previously undetected by the approaches tested in their study.
We report two interesting and motivating examples:

\noindent
\bcircle{1} \texttt{SecretLoc} identified new types of secrets (such as OpenAI API keys, GitHub tokens, and RSA private keys). For example, an OpenAI API key had previously gone undetected.. 
This occurred for two main reasons: (i) the regex-based detection method tested in~\cite{meli2019bad} did not account for the specific structure of OpenAI keys, and (ii) the OpenAI API methods were not incorporated into LeakScope~\cite{zuo2019does}. This illustrates how LLM-based approaches can uncover previously unseen secrets without requiring frequent updates to detection tools or predefined key formats.

\noindent
\bcircle{2} Existing regex-based tools are generally effective at detecting Google API keys due to their well-defined structure (i.e., \texttt{AIza[0-9A-Za-z-\_]{35}}, as shown by Li et al.~\cite{li2024automaticallydetectingcheckedinsecrets}). However, these tools can fail when keys are slightly modified or obfuscated—for example, by simple Base64 encoding. In our experiments, we encountered a Base64 string "QUl6YV*******..." (which decodes to "AIza**...") in an app that was completely missed by the regex-based system. In contrast, our LLM-based approach successfully identified the key, demonstrating its ability to detect API keys even when they deviate from known formats.

On the benchmark of Li et al.~\cite{li2024automaticallydetectingcheckedinsecrets}, we evaluated \textsc{SecretLoc} to determine to what extent our approach rediscovers previously confirmed secrets and identifies additional ones missed by existing tools. \textsc{SecretLoc} successfully detected 93\% of the secrets reported in the benchmark and uncovered \num{3671} additional ones, many of which were later validated through regex checks and manual inspection. Beyond this retrospective evaluation, we applied our approach to newly crawled apps from Google Play and identified clear-text secrets in the wild. Across a set of 5000 apps, we detected secrets in 2124 apps (42.5\%), several of which were confirmed and remediated by developers after we contacted them. We are continuing to monitor developer responses and provide assistance for proper remediation where needed.

Our insights show that LLMs can detect previously unseen or obfuscated secrets by leveraging both string structures and surrounding context, without relying on predefined patterns or training data. This capability, combined with the widespread presence of hardcoded secrets in mobile apps, underscores their potential for defensive use while highlighting the urgent need for proper secret management to prevent real-world attacks.

\noindent
\textbf{Contributions.} This paper makes the following contributions:
\begin{itemize}
\item We propose \texttt{SecretLoc}, an LLM-based approach for detecting secrets in Android apps, capable of identifying both known and previously unseen secrets.
\item We perform a comprehensive evaluation on the dataset of \num{5135} Android apps analyzed by Li et al.~\cite{li2024automaticallydetectingcheckedinsecrets}, showing that our approach detects secrets missed by existing regex-, static-, and machine-learning-based methods.
\item We demonstrate that \texttt{SecretLoc} can uncover obfuscated or modified secrets (e.g., Base64-encoded API keys) that evade conventional detection tools.
\item We extend our evaluation to newly crawled Google Play apps, detecting and responsibly disclosing previously unknown hardcoded secrets, highlighting the ongoing risks in real-world applications.
\item We provide actionable insights into the security implications of LLM-based secret detection, emphasizing both the defensive potential and the corresponding attack surface if these methods are misused.
\end{itemize}

\noindent
\textbf{Data Availability.} 
We provide all relevant resources to support further research and reproducibility. However, the public repository contains only a limited set of example apps, for which the exposed secrets have already been revoked by their developers, and can be safely used to test our approach. Access to the full dataset may be granted to qualified researchers upon reasonable request. 
Our artifacts are available at: 
\begin{center}
   \url{https://github.com/MarcoAlecci/LLMforAndroidSecrets}
\end{center}

\section{Background}
\label{sec:background}

In this section, we briefly present the concepts of secrets in mobile apps and the fundamentals of Android app reverse engineering, which form the basis of our study.

\noindent
\textbf{Secrets.} 
Cloud-based services typically require clients to authenticate using \emph{secrets}, such as API keys, client IDs, access tokens, or passwords. These secrets identify the calling app, regulate access to resources, and enforce billing policies. If exposed, adversaries can misuse them to: 
\wcircle{1} gain unauthorized access to sensitive data, 
\wcircle{2} corrupt or manipulate resources, 
\wcircle{3} generate unexpected financial costs by invoking paid APIs, or (iv) exhaust invocation limits, leading to denial-of-service (DoS)~\cite{CWE798,li2024automaticallydetectingcheckedinsecrets}. 

It is essential to distinguish between app secrets and user-sensitive information. User data refers to private information owned by Android users, such as contacts, photos, or calendars. This information is typically targeted at runtime by malicious apps that collect and transmit it to external servers, representing a different security challenge ~\cite{wei2025far}.

Some secrets have a well-defined structure that can be easily recognized and detected using regex-based approaches—for example, Google API Keys: \texttt{AIza[0-9A-Za-z\-\_]{35}}. However, other services do not rely on predictable patterns. For instance, Twitter Client Secrets consist of 40–50 alphanumeric characters, which lack a strong structural signature. This implies that regex-based tools require constant updates to accommodate new secret formats and may fail when no clear pattern exists, as is the case with Twitter. This limitation opens an opportunity for LLMs, which can infer whether a string is a secret by analyzing its surrounding context.

\medskip
\noindent
\textbf{Android App Reverse Engineering.} 
Android applications are typically distributed as compiled APK files through the Google Play Store or third-party marketplaces. An APK is a zip-compressed binary archive containing all components required to install and run an app, such as compiled code, resources, and metadata. The application logic is compiled into Dalvik Executable (DEX) bytecode, stored in .dex files~\cite{android_runtime}, which makes the original source code inaccessible in its distributed form.

Despite this, numerous tools exist for reverse engineering APKs, allowing for the recovery of human-readable representations of the code and resources~\cite{dex2jar, androguard, MobSF, vallee2010soot, apktool}. In our approach, we rely on ApkTool~\cite{apktool} and Soot~\cite{vallee2010soot}.
ApkTool unpacks APK files by disassembling both the .dex code and app resources. It decompiles bytecode into smali, a low-level but human-readable representation of Dalvik bytecode~\cite{smali_github}. This is particularly useful for analyzing app behavior and inspecting resource files. One common target of reverse engineering is the strings.xml file, which frequently contains developer-defined constants, including, at times, hardcoded secrets accidentally left in production apps.
In contrast, Soot provides high-level intermediate representations of Java bytecode, with Jimple being the most widely used. Jimple offers a simplified, three-address code format that facilitates static program analysis. With Soot, researchers and analysts can construct control-flow graphs, call graphs, and other program representations that enable systematic exploration of an app’s behavior~\cite{vallee2010soot}.

While reverse engineering tools are indispensable for researchers and security analysts, they are equally accessible to malicious actors. This dual-use nature underscores the importance of properly protecting secrets within Android apps, as exposed credentials can be easily extracted by anyone with basic reverse engineering skills.
\section{Approach}
\label{sec:approach}

\begin{figure*}[htb]
\centering
\includegraphics[width=\linewidth]{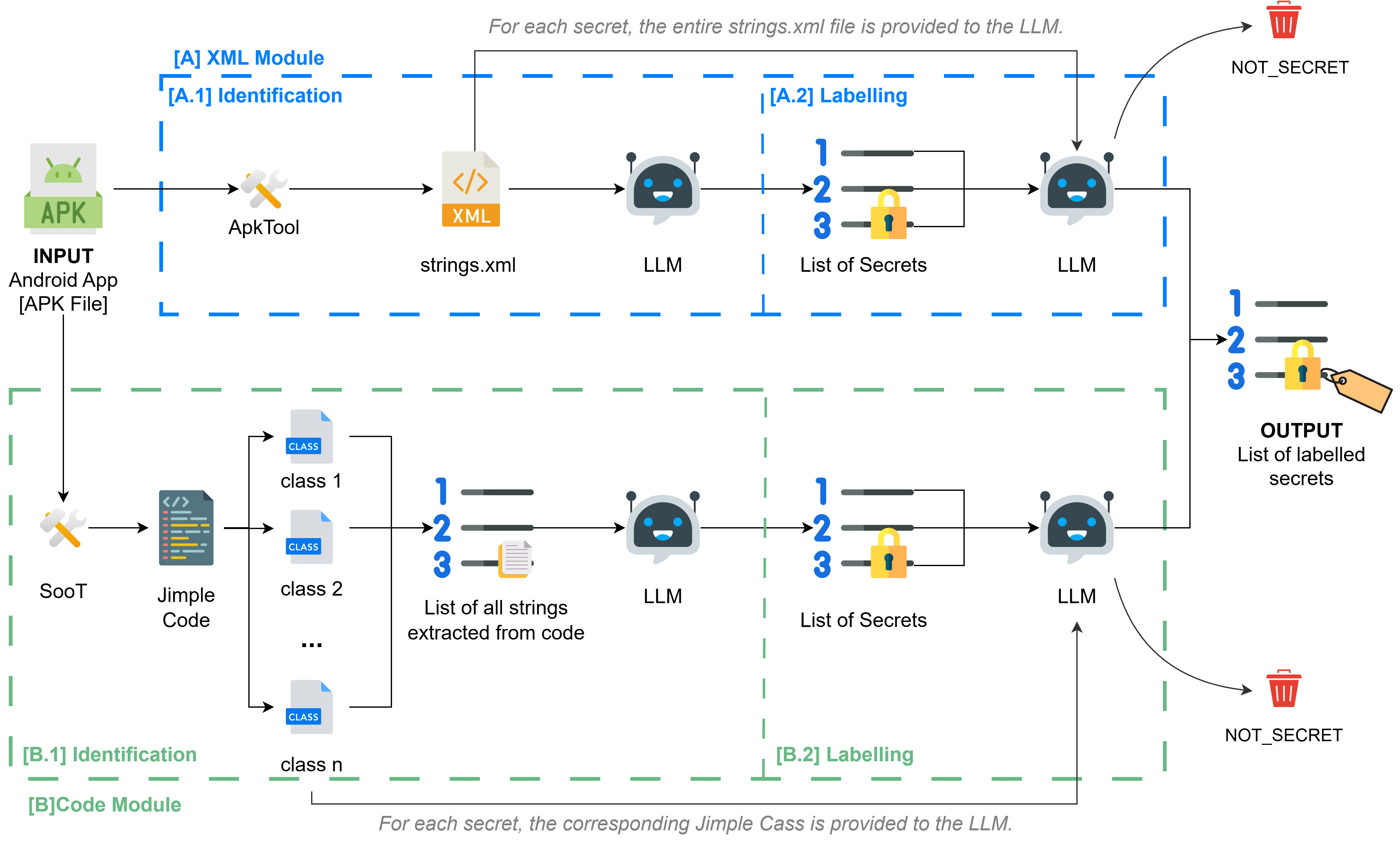}
\caption{Approach Overview.}
\label{fig:approachOverview}
\end{figure*}

The core idea behind \textsc{SecretLoc} is to leverage the contextual reasoning capabilities of LLMs to detect hardcoded secrets in Android applications. Unlike existing secret-detection tools, such as those based on regular expressions, static data-flow analysis, or supervised learning, our approach does not rely on any prior knowledge of secrets (e.g., regex patterns, API specifications for static analysis, or training data for machine learning).
Instead, we aim to evaluate whether an LLM can infer secrets directly from contextual and structural cues. This includes analyzing both the \texttt{strings.xml} resource file and the application’s source code, enabling the model to detect secrets defined as configuration constants (by interpreting XML element names and tags) or as hardcoded string literals within the implementation (by using surrounding code as context). 
An overview of the overall pipeline is shown in Figure~\ref{fig:approachOverview}. 

\textsc{SecretLoc} takes as input the APK file of the app under analysis and outputs a list of detected secrets, each accompanied by an automatically assigned label describing its type (e.g., \emph{Google API key}, \emph{OpenAI API key}, \emph{RSA private key}). The system consists of two main analysis modules:
\begin{itemize}[leftmargin=*]
    \item \textbf{[A] XML Module:} analyzes the \texttt{strings.xml} file, where developers often define constants that may contain embedded secrets.
    \item \textbf{[B] Code Module:} analyzes string constants extracted from the app’s bytecode, focusing on potentially sensitive values defined directly in the code.
\end{itemize}
Each module operates in two sequential phases: \textit{Identification} and \textit{Labeling}. In total, the pipeline thus comprises four phases: A1, A2, B1, and B2, which are described hereafter.

\noindent
\textbf{[A1]: XML Strings Identification.}
In this phase, \textsc{SecretLoc} extracts the app’s \texttt{strings.xml} file using \textit{ApkTool}~\cite{apktool}, which contains all textual resources (such as UI labels and constants) used by the Android application. Some developers, however, may store secrets such as credentials or API keys in this file. The entire XML file is provided to the LLM, which generates a list of candidate secrets. This list serves as input for the next phase.

\noindent
\textbf{[A2]: XML Strings Labeling.}
In this phase, each string identified in A1 is analyzed individually by the LLM. Unlike phase A1, where the entire XML file is examined at once, in A2 the LLM is asked to focus exclusively on the candidate string under analysis, using the rest of the \texttt{strings.xml} file only as supporting context. For each string, the LLM assigns a label indicating its secret type or service provider.
To reduce errors, the LLM can also assign a special label, \texttt{NOT-SECRET}, if the string is not actually a secret. The secrets receiving this label are then discarded. This allows correction of false positives from phase A1.
The output of A2 is a list of labeled secrets extracted from XML resources.

\noindent
\textbf{[B1]: Code String Identification.}
The code module focuses on code-level analysis. Using the \textit{Soot} framework, the APK is decompiled into its Jimple intermediate representation, from which all statically defined string constants are extracted. For each string, we keep track of the class from which it was extracted, as this context will be used in Phase B2.
The full list of extracted strings is provided to the LLM to determine potential secrets, without providing any additional context, with all strings analyzed together (i.e., not individually). This design choice is discussed in Section~\ref{sec:considerationsB1}.

\noindent
\textbf{[B2]: Code String Labeling.}
Each candidate secret identified in Phase B1 is then analyzed individually by the LLM for labeling. Along with the candidate string, the entire Jimple class from which it was extracted is provided as context, allowing the model to analyze the textual and semantic characteristics of each candidate and assign an appropriate label. As in the XML module, the LLM can assign the special \texttt{NOT-SECRET} label if the string is not actually a secret, allowing false positives from phase B1 to be discarded.

\subsection{Considerations about Phase [B1]}
\label{sec:considerationsB1}
In Phase B2, each string is accompanied by the Jimple class from which it was extracted, while this is not the case for Phase B1, where all strings are sent together without code context.
The main reason for this choice is scalability, as well as resource and cost efficiency. While it is feasible to analyze individually only a few candidate secrets (usually no more than 10) in Phase B2, analyzing individually all the extracted strings in Phase B1 would require sending thousands of individual requests (up to 10k for some apps), each including the entire Jimple class as context, which would dramatically increase computational cost and API usage. 
Therefore, in our design, Phase B1 serves more as a preprocessing step, allowing the actual identification to occur in Phase B2.

To assess the potential impact of this design choice on performance, in RQ4 we explore a variant of \textsc{SecretLoc} (on a small subset of our dataset) where each string extracted in Phase B1 is actually analyzed individually using the corresponding Jimple class as code context. This allows evaluation of the trade-off between accuracy and cost.

\subsection{Implementation Details}
\textsc{SecretLoc} is implemented using a combination of Java and Python. \textit{ApkTool} is employed to unpack APKs and extract XML resources, while \textit{Soot} is used for bytecode analysis and string extraction from decompiled Jimple code. 

For the LLM component, we rely primarily on \texttt{gpt-4o-mini} by OpenAI, selected for its balance between accuracy, cost, and inference speed. The impact of the underlying LLM model choice is evaluated later in Section~\ref{sec:rq3} (RQ3).

Overall, the modular and scalable design of \textsc{SecretLoc} enables efficient analysis of thousands of Android apps while remaining effective in identifying hardcoded secrets.
\section{Evaluation}
\label{sec:evaluation}

In this section, we evaluate our approach, \textsc{SecretLoc}, addressing the following research questions (RQs). The first two RQs focus on evaluating the performance of \textsc{SecretLoc}:
\begin{itemize}[leftmargin=*]
    \item \textbf{RQ1:} How does \textsc{SecretLoc} compare with existing secret detection tools?
    \item \textbf{RQ2:} How accurately does \textsc{SecretLoc} assign labels or categories to detected secrets?
\end{itemize}

The next two RQs aim to motivate our design choices through experiments:
\begin{itemize}[leftmargin=*]
    \item \textbf{RQ3:} How does the choice of the underlying LLM affect \textsc{SecretLoc}'s performance?
    \item \textbf{RQ4:} How does adding code context in the first phase affect \textsc{SecretLoc}'s performance?
\end{itemize}

Finally, we evaluate \textsc{SecretLoc} in a real-world scenario:
\begin{itemize}[leftmargin=*]
    \item \textbf{RQ5:} How does \textsc{SecretLoc} perform on a set of real-world apps from Google Play?
\end{itemize}

\noindent
\textbf{Datasets.}
To answer the first three RQs, we rely on the current state-of-the-art benchmark dataset of checked-in secrets in Android apps, created in a prior study by Li et al. and kindly released to us upon request~\cite{li2024automaticallydetectingcheckedinsecrets}. This dataset consists of \num{5135} APKs collected from Google Play, specifically the top 200 Android apps from each category. A complete description of the collection process is provided in their paper. In their study, the authors tested three secret detection tools and confirmed the presence of \num{2142} valid checked-in secrets across \num{2115} apps (around 41\% of all apps). Among these, \num{1836} secrets were detected by LeakScope~\cite{zuo2019does}, and  \num{666} by the regex-based approach (Three-Layer Filter) from Meli et al.~\cite{meli2019bad}, while PassFinder~\cite{feng2022automated} did not detect any valid secrets. Table~\ref{tab:checked-in-secrets} summarizes the types of confirmed secrets in the dataset that was shared with us. 

\begin{table}[ht]
    \centering
    \begin{adjustbox}{width=0.6\linewidth,center}
    \begin{tabular}{@{}l|cc@{}}
    \toprule
    \textbf{Service} & \textbf{Regex-based} & \textbf{LeakScope} \\ \midrule
    Google           & 664                  & 1816               \\
    Twitter          & 0                    & 10                 \\
    Facebook         & 0                    & 10                 \\
    Twilio           & 2                    & 0                  \\ \midrule
    \textbf{Total}   & 666                  & 1836               \\ \midrule
    \textbf{Union}   & \multicolumn{2}{c}{2142}                  \\ \bottomrule
    \end{tabular}
    \end{adjustbox}
    \caption{Distribution of confirmed checked-in secrets in the benchmark from~\cite{li2024automaticallydetectingcheckedinsecrets}.}
    \label{tab:checked-in-secrets}
\end{table}

To answer RQ4, we created a new dataset of real-world apps to further evaluate the capabilities of \textsc{SecretLoc} in a practical scenario.  For this purpose, we collaborated with the maintainers of AndroZoo~\cite{androzoo2016, alecci2024androzoo}, the largest publicly available collection of Android apps for research, which also provides metadata, including version codes. We requested a random sample of \num{5000} freshly collected apps from Google Play, ensuring that the apps were retrieved immediately after crawling, so that our analysis reflects the current state of the store. This dataset allows us to test \textsc{SecretLoc} on real-world applications and assess its ability to detect hardcoded secrets that may not be captured in existing benchmarks, providing a complementary evaluation to the benchmark dataset described above. 
For ethical reasons, we will release the full list of apps only upon request to certified researchers.

\subsection{RQ1: Comparison with Existing Tools}

\begin{figure}[htb]
    \centering
    \includegraphics[width=0.75\columnwidth]{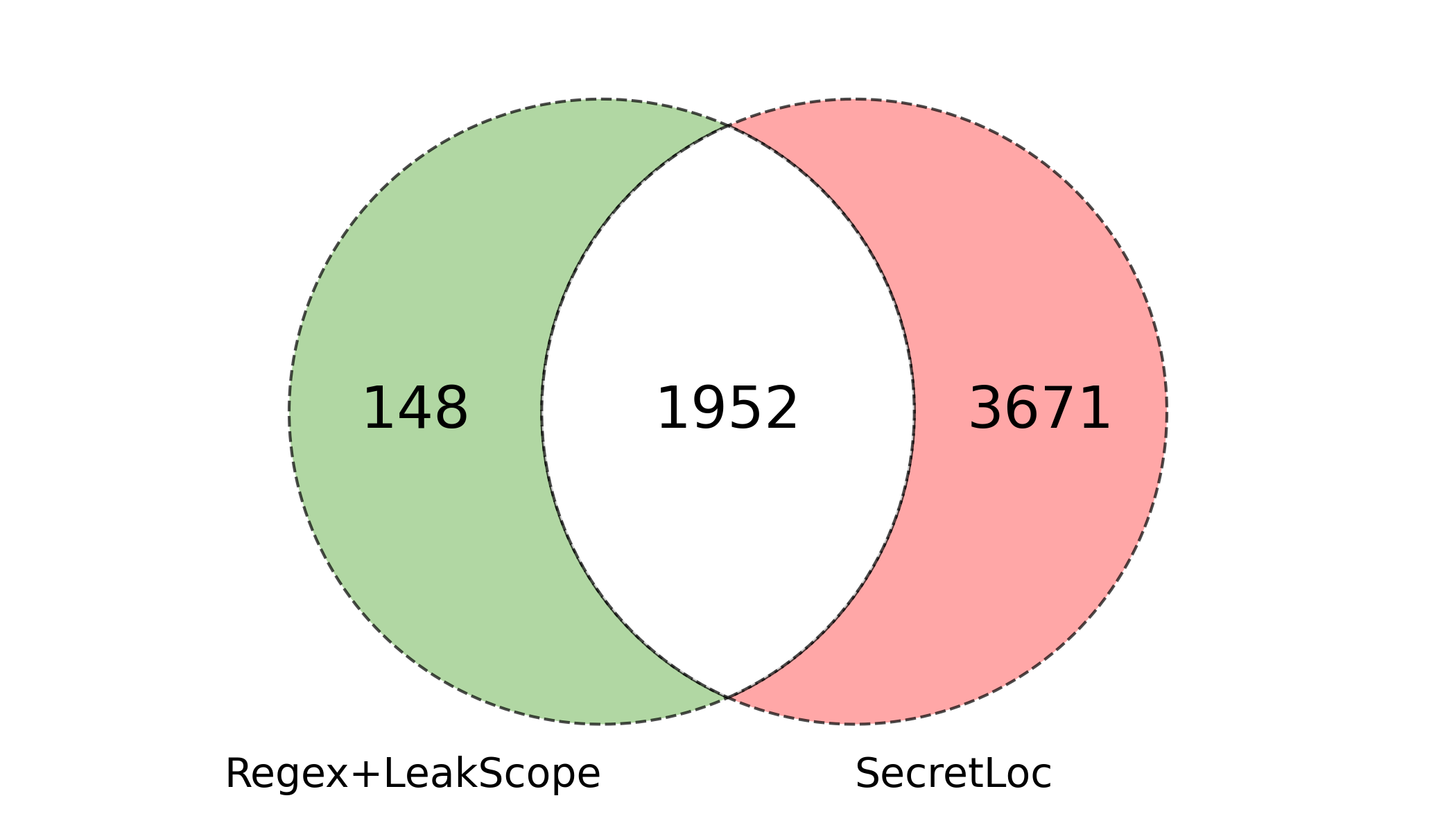}
    \caption{Comparison between existing tools and \textsc{SecretLoc}.}
    \label{fig:RQ1_vennDiagram}
\end{figure}

To address the first question, we tested \textsc{SecretLoc} on the \num{2115} apps from Li et al. ~\cite{li2024automaticallydetectingcheckedinsecrets} where secrets found by regex-based approaches or LeakScope have been manually confirmed by the authors. During our analysis, we were unable to analyze 56 apps due to issues with either Soot or ApkTool and excluded these from the analysis. The results are summarized in Figure~\ref{fig:RQ1_vennDiagram}. 

As observed, the vast majority of secrets previously identified by existing tools were also detected by \textsc{SecretLoc}, accounting for 93\% of all secrets reported in the benchmark. Notably, \textsc{SecretLoc} also discovered a large number of additional secrets, \num{3671}, that were missed by existing tools.
Given the large number of these “extra” potential secrets found by \textsc{SecretLoc} (which would make manual analysis extremely challenging and time-consuming), we employed two validation strategies to determine whether these newly discovered secrets were valid or false positives: \bcircle{1} verifying the findings using known regular expressions from Li et al.’s study~\cite{li2024automaticallydetectingcheckedinsecrets}, and \bcircle{2} manually inspecting a statistically significant random sample of results.

\noindent
\textbf{\bcircle{1} Validation Using Regular Expressions.}
Our first validation step consisted of applying the same regular expressions defined in Li et al.’s study~\cite{li2024automaticallydetectingcheckedinsecrets}. It is important to emphasize that this step was not intended to discover new secrets using regex (since our comparison includes a regex-based detector), but solely to confirm whether \textsc{SecretLoc}’s findings correspond to known secret formats. Accordingly, we restricted our analysis to regexes with well-defined and reliable patterns (e.g., Google and AWS keys) and excluded those with poorly defined structures, such as Twitter and Facebook secrets, as also noted by Li et al~\cite{li2024automaticallydetectingcheckedinsecrets}.
Table~\ref{table:extraSecrets} reports the results obtained through regex verification.

\begin{table}[h!]
\centering
\begin{adjustbox}{max width=\textwidth}
\begin{tabular}{l c}
\toprule
\textbf{Service} & \textbf{Count} \\
\midrule
Google API Key& 500 \\
Google\_OAuth & 1072 \\
Picatic & 0 \\
Stripe Standard API Key & 2 \\
Stripe Restricted API Key & 0 \\
Square Access Token & 0 \\
Square OAuth Secret & 0 \\
PayPal Braintree & 0 \\
Amazon MWS & 0 \\
Twilio & 0 \\
MailGun & 0 \\
MailChimp & 0 \\
Amazon AWS Access Key ID & 2 \\
\bottomrule
\end{tabular}
\end{adjustbox}
\caption{Additional secrets detected by \textsc{SecretLoc} confirmed through existing regex patterns.}
\label{table:extraSecrets}
\end{table}

As shown in Table~\ref{table:extraSecrets}, several of the additional secrets detected by \textsc{SecretLoc} were successfully confirmed using existing regex patterns, particularly for Google-related credentials. In total, we verified 1576 valid secrets (500 Google API keys, 1072 Google OAuth tokens, and a few from other services such as Stripe and AWS). These results indicate that a substantial portion of the “extra” findings correspond to genuine secrets that were previously overlooked by existing tools. However, regex-based confirmation inherently covers only secrets following known formats; thus, to further assess novel or less structured findings, we proceeded with manual inspection.

\noindent
\textbf{\bcircle{2} Manual Inspection.}
To complement automated validation, we performed a manual inspection on a statistically significant random sample of 94 secrets, selected using a 95\% confidence level and a 10\% margin of error from the set of secrets detected exclusively by \textsc{SecretLoc}. 
During this process, we analyzed secrets strictly at the code level to understand how they were used within the app (e.g., identifying their purpose or location in the logic). 
We did \emph{not} attempt to use any of the detected API keys, tokens, or passwords in external systems, thereby ensuring that no harm or unauthorized access could occur to developers or third-party services.

Our findings can be summarized as follows:
\begin{enumerate}[leftmargin=*]
    \item Out of the 94 inspected samples, 76 (81\%) were confirmed as valid secrets.  
    The majority of valid cases correspond to API keys and tokens for specific third-party services. However, among the inspected secrets, we also identified new types not covered by the benchmark, such as OpenAI API keys, Razorpay keys, LeanPlum, Kakao, Mapbox, and others. Some are extremely sensitive, such as a GitHub Personal Access Token we detected in one app. Moreover, we detected other types of secrets not specifically tied to a service, including RSA private keys, JWT tokens, database passwords/credentials, and even an Android keystore password used for signing APKs.  
    Among the false positives, most cases involved secrets that are actually public, such as Razorpay test API keys (used for simulated payments), Mapbox public keys, RSA public keys, and public identifiers like Facebook App ID or Google App ID. The remaining false positives were mostly cases where the LLM misclassified variable names as secrets—for example, labeling a variable named \texttt{"password"} as a real password, or \texttt{"access\_token"} as a real access token.  
    While these cases are not actual secrets, they can still lead to interesting findings. For instance, during inspection of the code containing the misclassified \texttt{"password"} variable, we discovered multiple issues, including SQL injection vulnerabilities, plaintext password storage, and real hardcoded credentials in the \texttt{resetDatabase()} method (where the variable appeared). These observations regarding false positives are discussed further in Section X.

    \item Furthermore, since several valid cases corresponded to API keys and tokens with well-defined, recognizable structures (e.g., OpenAI API keys, Razorpay, etc.), we decided to use regex again as a confirmation tool (i.e., we are only checking the secrets we detected that existing tools did not detect, using regex for confirmation rather than to discover new secrets). Table~\ref{table:newSecrets} summarizes these additional categories.

    \begin{table}[h!]
    \centering
    \begin{adjustbox}{max width=\textwidth}
    \begin{tabular}{l c}
    \toprule
    \textbf{Service} & \textbf{Count} \\
    \midrule
    OpenAI API Key              & 3  \\
    Razorpay (Live) API Key     & 6  \\
    RSA Private Key             & 2  \\
    JWT Token                   & 20 \\
    LeanPlum API Key            & 4  \\
    Kakao API Key               & 6  \\
    Mapbox API Key              & 4  \\
    \bottomrule
    \end{tabular}
    \end{adjustbox}
    \caption{New types of secrets detected by \textsc{SecretLoc}.}
    \label{table:newSecrets}
    \end{table}

    \item An especially interesting case involved a Google API key that was Base64-encoded, appearing in the app as the string \texttt{"QUl6YV*******..."}, which decodes to \texttt{"AIza**..."}. Such simple encoding can completely bypass standard regex detectors. In contrast, \textsc{SecretLoc} correctly identified it as a potential secret based on contextual clues. By decoding all Base64 strings found by SecretLoc and reapplying the regex-based verification, we confirmed 5 additional Google API keys that had been hidden using this technique.
    
\end{enumerate}

These results highlight how LLMs can detect secrets without any prior knowledge of the pattern, overcoming the limitations of tools such as LeakScope and regex-based approaches.

\subsubsection{Analysis of Previously “Clean” Apps}

In their study, Li et al. found and confirmed secrets in only \num{2115} out of \num{5135} apps, which we indeed used as our benchmark. However, since \textsc{SecretLoc} detected substantially more secrets overall, we extended the analysis to the remaining 3020 apps previously classified as clean, i.e., those not containing any secrets.
Our approach identified secrets in \num{2170} of these apps. We then verified these secrets using both the original and extended regex patterns (defined after the first manual inspection), confirmed a large number of valid secrets, as summarized in Tables~\ref{tab:RQ1_bis_standard} and~\ref{tab:RQ1_bis_extra}.

\begin{table}[h!]
\centering
\begin{adjustbox}{max width=\textwidth}
\begin{tabular}{l c}
\toprule
\textbf{Service} & \textbf{Count} \\
\midrule
Google & 2050 \\
Google\_OAuth & 1043 \\
Picatic & 0 \\
Stripe Standard API Key & 0 \\
Stripe Restricted API Key & 0 \\
Square Access Token & 0 \\
Square OAuth Secret & 0 \\
PayPal Braintree & 1 \\
Amazon MWS & 0 \\
Twilio & 0 \\
MailGun & 0 \\
MailChimp & 0 \\
Amazon AWS Access Key ID & 1 \\
\bottomrule
\end{tabular}
\end{adjustbox}
\caption{Standard secrets detected in previously “clean” apps.}
\label{tab:RQ1_bis_standard}
\end{table}

\begin{table}[h!]
\centering
\begin{adjustbox}{max width=\textwidth}
\begin{tabular}{l c}
\toprule
\textbf{Service} & \textbf{Count} \\
\midrule
OpenAI API Key              & 2 \\
Razorpay API Key LIVE       & 7 \\
RSA Private Key             & 0 \\
JWT Token                   & 17 \\
LeanPlum API Key            & 2 \\
Kakao API Key               & 2 \\
Mapbox API Key              & 4 \\
\bottomrule
\end{tabular}
\end{adjustbox}
\caption{New categories of secrets detected in previously “clean” apps.}
\label{tab:RQ1_bis_extra}
\end{table}

Additionally, decoding Base64 strings revealed 3 more Google API keys that had previously gone undetected by existing tools. This implies that \textsc{SecretLoc} successfully identified secrets in numerous apps where previous tools had detected none.

\begin{answer}
\textbf{Answer to RQ1:}  
\textsc{SecretLoc} rediscovered 93\% of the secrets identified by existing tools and detected additional ones, many of which were validated as genuine through regex checks and manual inspection. It also uncovered new categories (e.g., OpenAI, Razorpay, JWT) and obfuscated secrets missed by prior approaches, demonstrating its superior coverage and generalization capabilities.
\end{answer}

%%%%%%%%%%%%%%%%%%%%%%%%%%%%%%%%%%%
\subsection{RQ2: Accuracy of Assigned Labels}

\textsc{SecretLoc} also returns a label for each detected secret, providing an indication of its type or the service it belongs to. To evaluate whether these labels are accurate, we analyzed the same benchmark set of \num{2115} apps from Li et al.~\cite{li2024automaticallydetectingcheckedinsecrets}. We employed two complementary validation strategies, similar to those used in RQ1: \bcircle{1} verifying the labels of known secrets from Li et al.’s benchmark study, and \bcircle{2} manually inspecting a statistically significant random sample of results.

\begin{figure*}[htb]
\centering
\includegraphics[width=\linewidth]{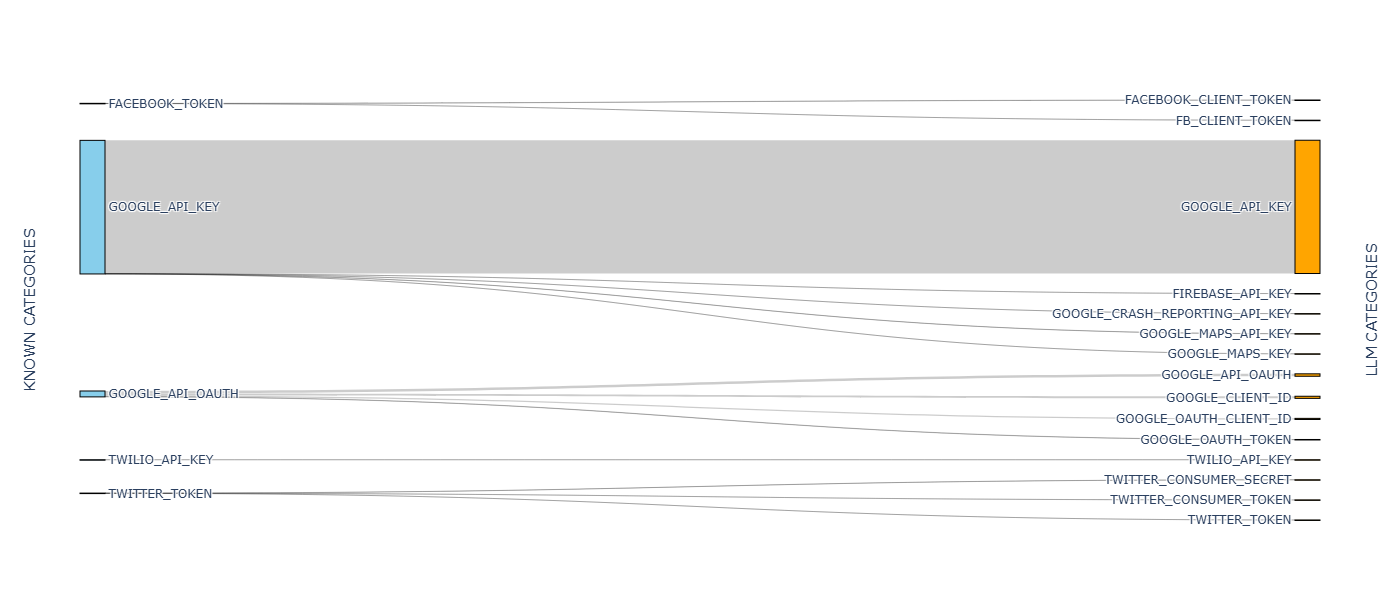}
\caption{Mapping between benchmark categories (derived from regex/API matches in Li et al.~\cite{li2024automaticallydetectingcheckedinsecrets}) and the labels assigned by \textsc{SecretLoc}.}
\label{fig:RQ2_sankeyDiagram}
\end{figure*}

\noindent
\textbf{\bcircle{1} Validation over known secrets.}
In Li et al.’s dataset, each secret is associated with a \emph{category} derived from the matching rule that identified it (e.g., a regex or predefined API signature). These categories are not explicitly assigned by the tools themselves, but rather reflect the origin of the matching pattern (e.g., “Google API Key” or “AWS Access Key ID”). 
For this analysis, we compared those pattern-derived categories against the semantic labels generated by \textsc{SecretLoc}. Specifically, we examined whether the LLM-assigned labels were consistent with or refined the reference categories. The comparison results are summarized in Figure~\ref{fig:RQ2_sankeyDiagram}.

As observed, the categories produced by \textsc{SecretLoc} were highly consistent with those inferred from the benchmark. In several cases, the LLM generated more fine-grained or context-aware labels than the original pattern-based categories. For example, while regex-based methods could only indicate that a credential belonged to the broader “Google Services” group, \textsc{SecretLoc} could further specify the exact service, such as \emph{YouTube}, \emph{Google Maps}, or \emph{Google Drive}. 
Without explicit normalization constraints, however, the LLM occasionally produced semantically equivalent label variants (e.g., slightly different formulations for Google OAuth tokens) as it can be observed for Facebook or Twitter tokens. Although these variations do not affect correctness, they suggest an opportunity for future work to unify label terminology.

\noindent
\textbf{\bcircle{2} Manual Inspection.}
To complement the benchmark-based validation, we manually inspected a random sample of 94 detected secrets (95\% confidence level, 10\% margin of error) to assess the accuracy and granularity of the assigned labels.

Out of the 94 samples, we observed the following results (summarized in Table~\ref{tab:RQ2_manual_inspection}):
\begin{itemize}
    \item \textbf{15 cases (16\%)} were false positives, i.e., non-secret strings mistakenly labeled as secrets. For example, a Razorpay test key was incorrectly labeled as \texttt{RAZORPAY\_API\_KEY}.
    \item \textbf{3 cases (3\%)} were confirmed secrets but had incorrect labels (e.g., a Google OAuth token mislabeled as a Google API key).
    \item \textbf{2 cases (2\%)} were valid secrets but assigned overly generic labels. For instance, a StableDiffusion API key was labeled simply as \texttt{API\_KEY}, while a Baron Weather API credential was labeled as \texttt{PRIVATE\_KEY}.
    \item The remaining \textbf{74 cases (79\%)} were correctly identified and accurately labeled. As also illustrated in Figure~\ref{fig:RQ2_sankeyDiagram}, some of these correct cases involved synonymous label variants (e.g., multiple naming forms for the same service). Such variations are semantically equivalent and could be trivially consolidated post hoc without affecting correctness.
\end{itemize}

\begin{table}[htb]
\centering
\caption{Results of manual inspection on 94 randomly selected secrets detected by \textsc{SecretLoc}.}
\label{tab:RQ2_manual_inspection}
\begin{tabular}{lcc}
\toprule
\textbf{Category} & \textbf{Count} & \textbf{Percentage} \\
\midrule
Correct secret \& correct label & 74 & 79\% \\
Valid secret, wrong label & 3 & 3\% \\
Valid secret, overly generic label & 2 & 2\% \\
False positive (not a secret) & 15 & 16\% \\
\midrule
\textbf{Total} & \textbf{94} & \textbf{100\%} \\
\bottomrule
\end{tabular}
\end{table}

Overall, the vast majority of labels were accurate, with most errors corresponding to false positives (i.e., the label was incorrect because the string was not actually a secret), and a few cases involving overly generic or misclassified labels.

\begin{answer}
\textbf{Answer to RQ2:}
\textsc{SecretLoc} accurately assigns meaningful labels to detected secrets. For all known secrets in the benchmark dataset, the LLM-generated labels were consistent with or more specific than the pattern-based categories produced by existing detection rules. Manual validation further confirmed that 79\% of inspected labels were correct, with the remaining cases involving minor mislabeling, overly generic labels, or false positives. These results demonstrate that \textsc{SecretLoc} can not only identify secrets reliably but also infer their semantic types with high precision.
\end{answer}

%%%%%%%%%%%%%%%%%%%%%%%%%%%%%%%%%%
\subsection{RQ3: Impact of the Underlying LLM Choice}
\label{sec:rq3}

\begin{figure*}[t]
    \centering
    \subfloat[\textbf{Detection Recall.} 
    Percentage of known secrets correctly identified across different LLMs using the benchmark dataset.]{
        \includegraphics[width=0.46\textwidth]{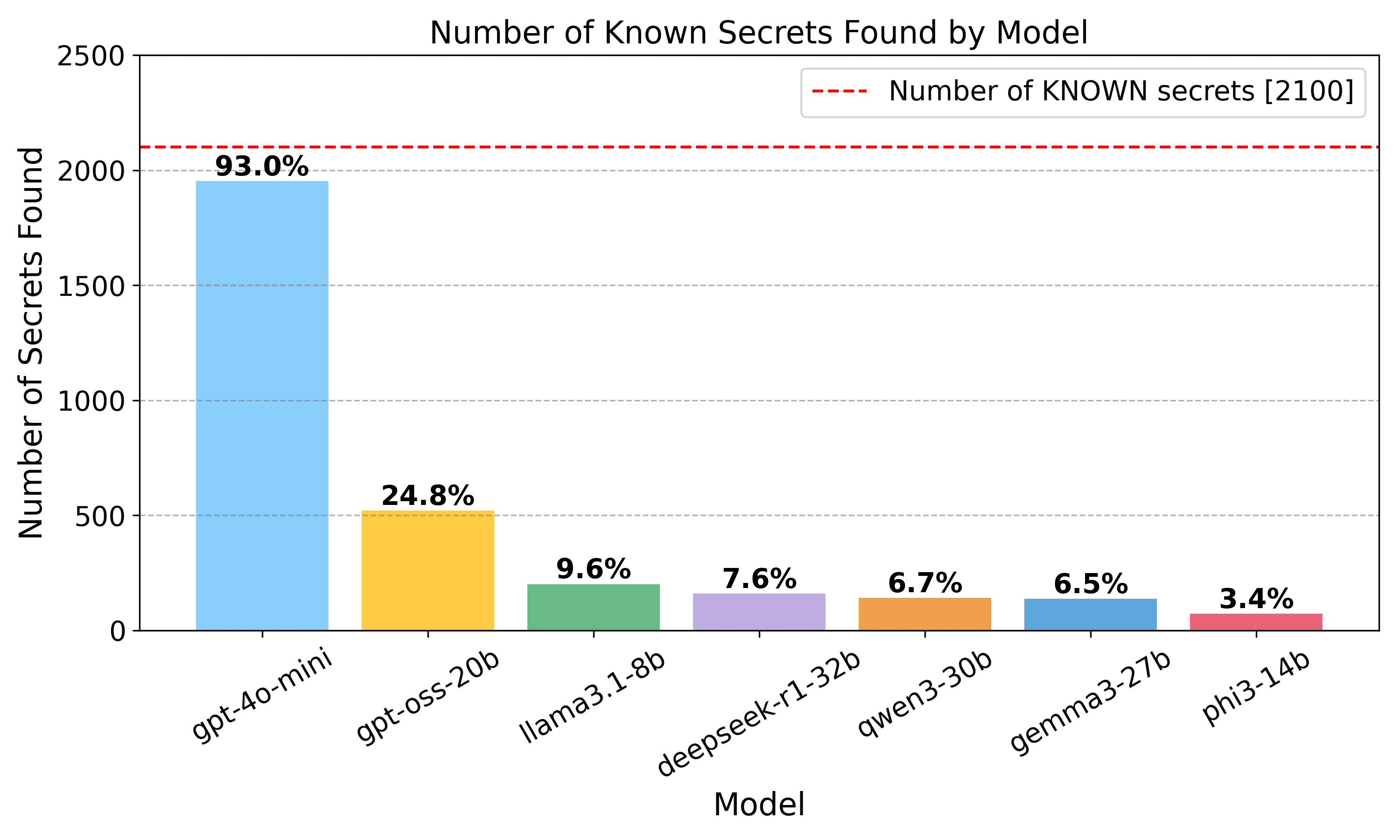}
        \label{fig:RQ3_known}
    }
    \hfill
    \subfloat[\textbf{Extra Secrets Confirmed.} 
    Number of additional valid secrets detected beyond the benchmark, verified through regex-based validation.]{
        \includegraphics[width=0.46\textwidth]{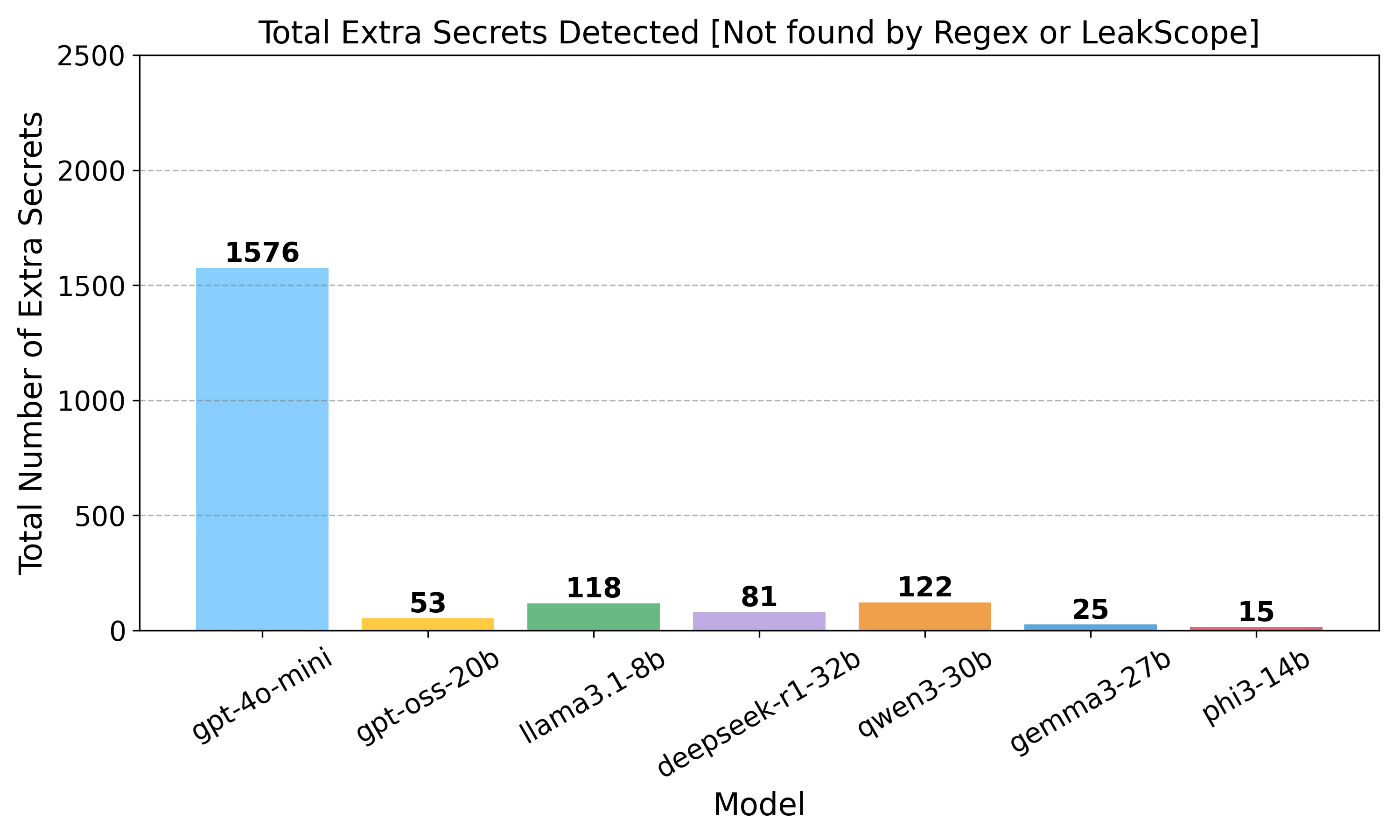}
        \label{fig:RQ3_extra}
    }
    \caption{Comparison of the performance of different LLMs integrated in \texttt{SecretLoc}.}
    \label{fig:RQ3}
\end{figure*}

While developing \texttt{SecretLoc}, we faced several key design decisions, one of the most important being the choice of the underlying large language model (LLM). We initially adopted a relatively lightweight and cost-efficient model, \texttt{gpt-4o-mini}, but we wanted to evaluate whether open-source LLMs could achieve comparable performance in detecting hardcoded secrets.
To this end, we conducted experiments on the same \num{2115} apps from the benchmark dataset of Li et al.~\cite{li2024automaticallydetectingcheckedinsecrets}. We executed all open-source models locally on an NVIDIA RTX5000 ADA GPU using the \texttt{ollama} deployment framework.  For each family of open-source models (e.g., LLaMA, Gemma, and GPT-OSS), we selected the largest model (in terms of parameters) that could be executed within our available hardware resources. 
We evaluated each model according to two metrics: \bcircle{1} the percentage of known secrets (from the benchmark) successfully identified by \texttt{SecretLoc} and \bcircle{2} the number of additional secrets detected beyond the benchmark, verified through existing regular expressions from Li et al.’s study~\cite{li2024automaticallydetectingcheckedinsecrets}.
The results are summarized in Figure~\ref{fig:RQ3}, which presents both the percentage of known secrets detected (Figure~\ref{fig:RQ3_known}) and the number of extra secrets confirmed via regex validation (Figure~\ref{fig:RQ3_extra}).

As shown in Figure~\ref{fig:RQ3}, the performance gap between \texttt{gpt-4o-mini} and the open-source models is substantial. The proprietary OpenAI model \texttt{gpt-4o-mini} correctly identified 93\% of the known secrets from the benchmark, while the best-performing open-source alternative, \texttt{gpt-oss-20b}, reached only 25\%. It is interesting to note that among all the open-source models, \texttt{gpt-oss-20b} performs on a different level compared to the others, which show similar performance among each other. A similar trend is observed when considering the number of additional secrets discovered. \texttt{gpt-4o-mini} detected \num{1576} additional secrets validated by existing regex patterns, whereas \texttt{gpt-oss-20b} found only 53 such cases.

Based on these results, we decided to rely on \texttt{gpt-4o-mini}. While it is true that other proprietary models, and even more powerful models from OpenAI itself, could be tested, we believe that \texttt{gpt-4o-mini} represents the best trade-off between performance and cost, with the analysis of an app costing less than \$0.01.
Since our primary goal is to evaluate the capabilities of large language models in this specific task, rather than benchmarking every available model, we consider the performance of \texttt{gpt-4o-mini} to be sufficiently strong to support our study and provide meaningful insights into the problem.

\begin{answer}
\textbf{Answer to RQ3:} 
The choice of the underlying LLM has a major impact on \texttt{SecretLoc}'s performance. 
\texttt{gpt-4o-mini} achieved the best results, detecting 93\% of known secrets and discovering more than 1500 additional valid ones. 
In contrast, the best open-source model (\texttt{gpt-oss-20b}) achieved only 25\% recall with significantly fewer validated detections. These results confirm that proprietary models currently offer superior contextual reasoning for secret detection.
\end{answer}

%%%%%%%%%%%%%%%%%%%%%%%%%%%%%%%%%%
\subsection{RQ4: Impact of Adding Code Context in the First Phase}
\label{sec:rq4}

As discussed previously in Section~\ref{sec:considerationsB1}, we made a specific design choice in \textsc{SecretLoc} to preserve scalability as well as resource and cost efficiency. In particular, during Phase~B1 (see Section~\ref{sec:approach}), all strings are analyzed collectively without incorporating their surrounding code context, which is used only later in Phase B2. In this research question, we explore a variant of \textsc{SecretLoc} in which each extracted string is analyzed individually, together with its corresponding Jimple class, thereby providing the model with contextual information already during the initial identification phase B1. The goal of this experiment is to investigate whether introducing such context at earlier stage can improve the detection rate and accuracy of secrets, while also quantifying the computational and financial overhead introduced by this design choice.

To avoid incurring prohibitive costs, caused by the substantially higher number of LLM requests required in Phase~B1, we evaluated this variant on a statistically significant random sample of 96 apps (95\% confidence level, 10\% margin of error) out of the \num{2115} benchmark apps used in previous RQs. As expected, both the average analysis time and the cost per app increased substantially. The performance comparison is summarized in Table~\ref{tab:rq4_time_cost}.

\begin{table}[h]
\centering
\caption{Comparison of performance metrics between the original \textsc{SecretLoc} and the contextual variant.}
\label{tab:rq4_time_cost}
\begin{adjustbox}{width=0.9\columnwidth}
\begin{tabular}{l|cc|l}
\hline
\multicolumn{1}{c|}{\textbf{Metric}} &
  \textbf{\begin{tabular}[c]{@{}c@{}}SecretLoc\\ {[}Original{]}\end{tabular}} &
  \textbf{\begin{tabular}[c]{@{}c@{}}SecretLoc \\ {[}Variant{]}\end{tabular}} &
  \textbf{Increase} \\ \hline
Avg. Time per app  &  43s     &  1320s  &  +2969\% \\
Avg. Cost per app  &  0.008\$ &  0.79\$ &  +9775\%
\end{tabular}
\end{adjustbox}
\end{table}

The significant overhead primarily stems from the large number of strings, and consequently, the high volume of requests, that must be processed individually in the contextual variant. For instance, one analyzed app contained \num{7761} strings, leading to a total analysis cost of 2.83\$ and a processing time of \num{7312} seconds. In contrast, the same app required only 56 seconds and 0.009\$ in the original configuration, with no difference in the number of detected secrets.

\begin{figure}[htb]
    \centering
    \includegraphics[width=0.75\columnwidth]{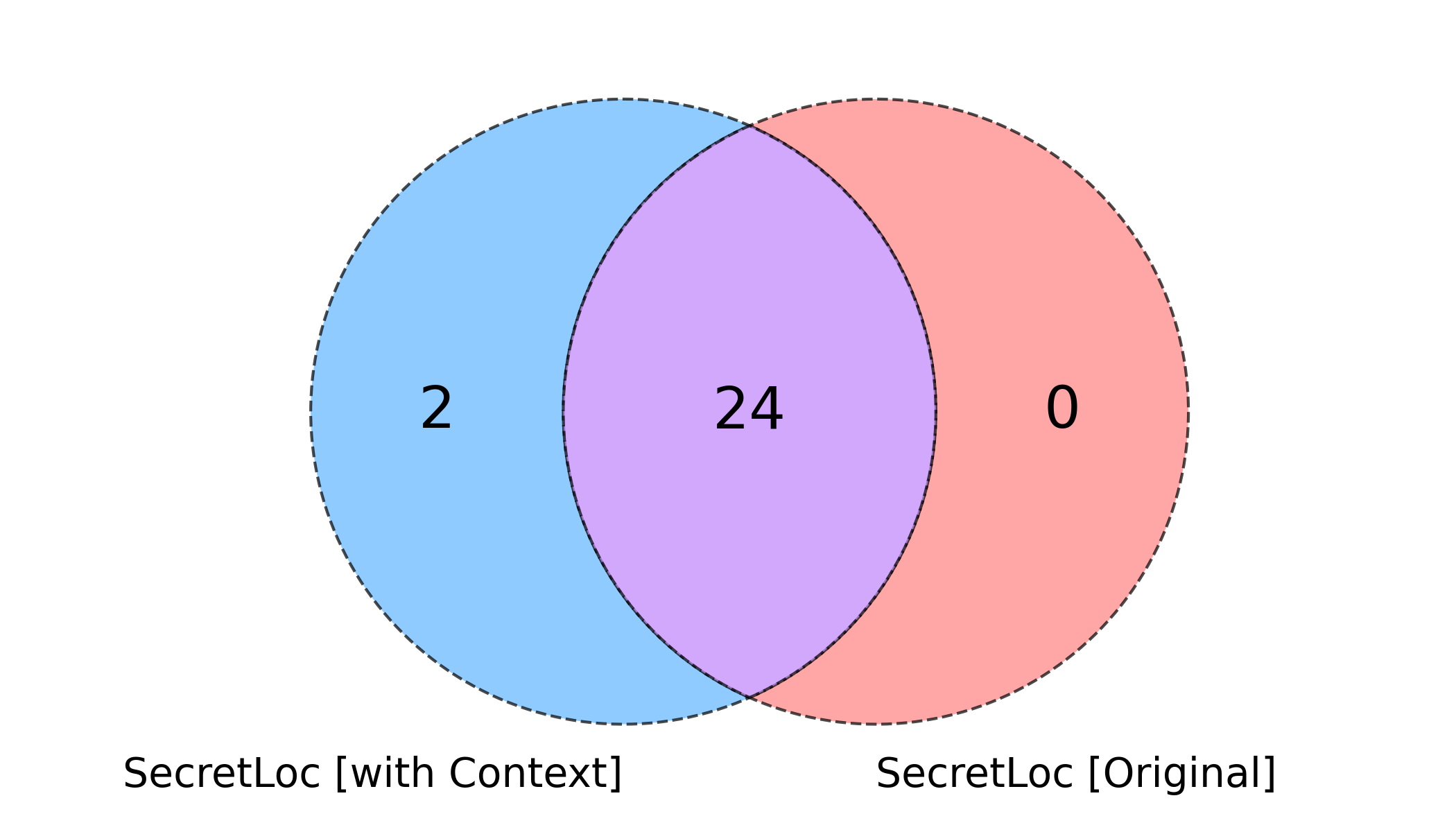}
    \caption{Overlap of secrets detected in code by the original and contextual variants of \textsc{SecretLoc}.}
    \label{fig:RQ4_vennDiagram}
\end{figure}

Although the overhead (in both time and cost) is substantial, we further examined whether the contextual variant improves the number of detected secrets, given that code context is already considered in Phase~B1. To this end, we focused exclusively on secrets located in code (i.e., excluding those found in XML resources) and compared the results between the two configurations. The comparison is illustrated in Figure~\ref{fig:RQ4_vennDiagram}.

As shown in the figure, the vast majority of secrets (92\%) were detected by both configurations, with only two additional secrets identified exclusively by the contextual variant. These were API keys for \texttt{YANDEX\_METRICA}—an analytics platform—and \texttt{BRAZE}, a customer engagement platform used for push notifications and email campaigns. Both keys resemble UUID patterns (8-4-4-4-12 hexadecimal format), which likely caused the original pre-filtering stage to mistakenly discard them as random UUIDs. In contrast, by analyzing each string together with its class, the contextual variant enabled the LLM to infer their true purpose from the surrounding code.

Overall, the benefits of adding early code context are confined to such isolated edge cases, while the associated overhead remains prohibitive for large-scale analyses.

\begin{answer}
\textbf{Answer to RQ4:}  
Although incorporating code context during the first phase can occasionally improve detection in specific cases, the considerable increase in cost and analysis time outweighs the marginal accuracy gains. Therefore, the original design of \textsc{SecretLoc} remains the most practical and cost-effective configuration for large-scale secret detection.
\end{answer}

%%%%%%%%%%%%%%%%%%%%%%%%%%%%%%%%%%%%%
\subsection{RQ5: Performance on Real-World Apps}

So far, our evaluation relied on the benchmark dataset created by Li et al.~\cite{li2024automaticallydetectingcheckedinsecrets}, which represents the state-of-the-art in secret detection for Android apps. In this research question, we aim to evaluate the performance of \textsc{SecretLoc} in a real-world scenario. To this end, we created a new dataset of \num{5000} apps freshly crawled from Google Play, with the help of the AndroZoo team~\cite{androzoo2016}.

When running \textsc{SecretLoc}, we detected at least one secret in \num{2124} apps out of \num{5000}, corresponding to 42.5\% of the dataset. The fact that these apps came directly from Google Play highlights the ongoing prevalence of hardcoded secrets. However, it is necessary to assess the nature of these secrets. We applied the same two validation strategies used in RQ1: \bcircle{1} verification using both known and newly defined regular expressions, and \bcircle{2} manual inspection of a statistically significant random sample.

\noindent
\bcircle{1} Validation Using Regular Expressions
We applied all previously defined regex patterns, including those extended in RQ1, to classify detected secrets. The results are summarized in Tables~\ref{tab:RQ5_standard} and~\ref{tab:RQ5_extra}.

\begin{table}[h!]
\centering
\begin{adjustbox}{max width=\textwidth}
\begin{tabular}{l c}
\toprule
\textbf{Service} & \textbf{Count} \\
\midrule
Google & 1684 \\
Google\_OAuth & 820 \\
Picatic & 0 \\
Stripe Standard API Key & 0 \\
Stripe Restricted API Key & 0 \\
Square Access Token & 0 \\
Square OAuth Secret & 0 \\
PayPal Braintree & 0 \\
Amazon MWS & 0 \\
Twilio & 0 \\
MailGun & 0 \\
MailChimp & 0 \\
Amazon AWS Access Key ID & 2 \\
\bottomrule
\end{tabular}
\end{adjustbox}
\caption{Confirmed standard secrets detected by \textsc{SecretLoc} in real-world Google Play apps using regex validation.}
\label{tab:RQ5_standard}
\end{table}

\begin{table}[h!]
\centering
\begin{adjustbox}{max width=\textwidth}
\begin{tabular}{l c}
\toprule
\textbf{Service} & \textbf{Count} \\
\midrule
OpenAI API Key & 0 \\
Razorpay API Key LIVE & 3 \\
Razorpay API Key TEST & 1 \\
RSA Private Key & 0 \\
JWT Token & 2 \\
\bottomrule
\end{tabular}
\end{adjustbox}
\caption{Additional confirmed secrets detected by \textsc{SecretLoc} in real-world Google Play apps using regex validation.}
\label{tab:RQ5_extra}
\end{table}

As observed, the majority of confirmed secrets are Google-related, consistent with the findings of Li et al.~\cite{li2024automaticallydetectingcheckedinsecrets}. Nevertheless, these results suggest that hardcoded secrets persist in apps available on Google Play.

\noindent
\bcircle{2} Manual Inspection
We performed a manual inspection on a statistically significant random sample of 94 detected secrets (95\% confidence level, 10\% margin of error). 
During this process, we analyzed (as for RQ1) the detected strings strictly at the code level to understand their usage within the app (e.g., purpose, context, or function) without attempting to use any of the keys or tokens in external systems, thereby ensuring no harm to developers or third-party services.  

Out of the 94 inspected samples, 74 (79\%) were confirmed as valid secrets. The remaining 20 (21\%) corresponded to false positives, whose nature was consistent with what we previously observed in RQ1 and RQ2—primarily public or test credentials (e.g., Mapbox public keys, Razorpay test keys) and benign identifiers misclassified due to variable naming (e.g., \texttt{"password"} variables). Among the confirmed secrets, we identified several new types of credentials, including \textbf{Klaviyo}, \textbf{Alibaba Cloud}, \textbf{HuggingFace}, and others.  
These findings further illustrate the diversity of secrets hardcoded in production apps and emphasize that secret exposure continues to affect both established and emerging service providers.

\subsubsection{Disclosure to Developers}
\label{sec:developers}

After validating detected secrets using regex and manual inspection, we disclosed the findings to app developers. Developer contact emails were retrieved from Google Play, and emails were sent for each app with a detected secret. For safety, we did not include the secret itself in the email body but notified the developers that a potentially hardcoded secret may exist in their app.
Out of the \num{2124} affected apps, we were able to retrieve developer email addresses for \num{1791} apps. Of these, 118 emails were returned as undeliverable. We received confirmations from 15 developers indicating that they are addressing or have already addressed the reported issue. We continue to receive responses and are actively monitoring developer feedback to provide clarification and guidance where needed to support proper remediation.

\begin{answer}
\textbf{Answer to RQ5:}
\textsc{SecretLoc} detected at least one hardcoded secret in \num{2124} out of \num{5000} real-world Google Play apps (42.5\%). Regex validation confirmed a large proportion of these secrets as genuine, primarily Google-related keys, while additional manual inspection revealed other secret types. The findings demonstrate that hardcoded secrets are still widespread in real-world apps. Responsible disclosure to developers was performed, resulting in confirmations that some issues were resolved.
\end{answer}
\section{Discussion}
\label{sec:discussion}

While \textsc{SecretLoc} demonstrates strong performance in detecting hardcoded secrets, several important considerations emerge from our evaluation.

\noindent
\textbf{False Positives and Candidate Prioritization}
Like any detection tool, \textsc{SecretLoc} is not perfect and generates false positives. However, these false positives are not meaningless: they can serve as valuable indicators for other potential vulnerabilities. For instance, during our manual analysis, a string labeled as a secret (e.g., "password") was not a secret itself but highlighted a Jimple class containing multiple issues, including SQL injection vulnerabilities, plaintext password storage, and real hardcoded credentials in the resetDatabase() method. This shows that even false positives can guide security analysts to relevant code regions, reducing the search space and prioritizing manual inspection.

\noindent
\textbf{Practicality for Defensive Use}
In practice, most apps contain a limited number of secrets—usually no more than five. This means that, even with some false positives, \textsc{SecretLoc} can effectively pinpoint candidate hardcoded secrets in a single app or a small set of apps. Consequently, the tool is highly practical for defensive purposes, allowing developers or security teams to focus their attention on a manageable subset of potentially sensitive values.

\noindent
\textbf{Efficiency and Scalability}
\textsc{SecretLoc} is straightforward to deploy and scales efficiently. On average, it requires approximately 43 seconds per app at a cost of around \$0.008 per app. This combination of speed and affordability makes it suitable for both individual app audits and larger-scale analyses of multiple applications.

\noindent
\textbf{Potential Risks and Dual-Use Concerns}
The simplicity and efficiency of the approach also imply potential misuse. An attacker could leverage \textsc{SecretLoc} to systematically analyze apps on platforms like Google Play, identifying hardcoded secrets for malicious purposes. This dual-use nature highlights the need for responsible handling and ethical disclosure when employing LLM-based detection tools.

\noindent
\textbf{Contextual Security Considerations}
Not all detected secrets are inherently dangerous. Some API keys, such as Firebase API keys, are generally considered safe to expose in code~\cite{firebase_api_keys}. However, their safety depends on proper configuration. If associated project IDs or authentication tokens are mismanaged or leaked, attackers could still exploit these secrets, as confirmed by Li et al.~\cite{li2024automaticallydetectingcheckedinsecrets}, who demonstrated that additional sensitive information could be extracted from seemingly harmless Firebase API keys. This reinforces the idea that the risk of exposed secrets is context-dependent and that even low-sensitivity credentials may become exploitable when combined with other data.

\noindent
\textbf{Summary of Implications}
In summary, \textsc{SecretLoc} provides a practical and scalable approach to secret detection, capable of uncovering both known and previously unseen secrets. While false positives exist, they often highlight code areas of interest and reduce manual effort. The method is effective for small-scale defensive audits but could also be abused if deployed at scale by attackers. Finally, the security impact of detected secrets depends not only on the secret itself but also on the surrounding context and service configuration, highlighting the need for better security practices.

\section{Limitations}
\label{sec:limitations}
In this section, we discuss the main limitations of our approach, focusing on both technical constraints and methodological challenges.

\noindent
\textbf{String Extraction.}
\textsc{SecretLoc} relies on two main sources of string data: values defined in \texttt{strings.xml} and hardcoded string constants extracted from the app code. This design allows us to capture the majority of embedded secrets, but it introduces two notable limitations.
\wcircle{1} \emph{Dynamic string construction.} Strings that are built at runtime—e.g., through \texttt{StringBuilder}, \texttt{append()}, or concatenation operations—may not be detected, as \textsc{SecretLoc} currently focuses on statically defined constant values. This is a common limitation shared by most static-analysis-based approaches.
\wcircle{2} \emph{String obfuscation.} Some apps employ code obfuscation techniques. While identifier renaming is relatively widespread, previous work (e.g., Dong et al.~\cite{dong2018understanding}) suggests that explicit string encryption remains uncommon in real-world Android apps (0\% of apps they analyzed from Google Play and 0.01\% of apps from third-party markets).
Unlike traditional regex-based methods, \textsc{SecretLoc} can successfully handle simple obfuscation schemes such as Base64 encoding (as demonstrated in Section~\ref{sec:evaluation}). However, detecting secrets hidden through advanced encryption remains significantly more challenging. Given the rarity of such strong encryption in typical apps, we consider this limitation acceptable within the current scope of our study.

\noindent
\textbf{LLM-Related Challenges.}
As an LLM-driven system, \textsc{SecretLoc} inherits several intrinsic limitations of large language models, including occasional hallucinations where the model produces incorrect or overconfident classifications. This means that false positives and false negatives may still occur. In principle, techniques such as improved prompt engineering, consensus-based prompting, or the use of more powerful LLMs could help reduce hallucinations and misclassifications. However, in this study, our primary objective was to evaluate the feasibility and baseline capabilities of LLMs for secret detection rather than to optimize performance through prompt or model engineering. We therefore leave a systematic exploration of such enhancements to future work.

\noindent
\textbf{Recall Estimation.}
Accurately measuring the true recall of \textsc{SecretLoc} is inherently challenging. Our recall metrics are computed with respect to the benchmark dataset provided by Li et al.~\cite{li2024automaticallydetectingcheckedinsecrets}. However, since \textsc{SecretLoc} discovered many additional secrets beyond those included in the benchmark, determining whether these represent true positives or false positives is difficult. The lack of ground truth for newly detected secrets makes it impractical to compute absolute recall. To mitigate this, we employed a two-step validation process, combining regex-based verification and manual review, to estimate detection quality, though some uncertainty inevitably remains.
\section{Related Work}
\label{sec:relatedwork}

In this section, we briefly review prior work on secret detection both in Android apps and generic code repositories, as well as security in Android.

\noindent
\textbf{Secrets Detection in Android Apps.}
Our study builds upon the dataset released by Li et al.~\cite{li2024automaticallydetectingcheckedinsecrets}, which represents the current state of the art in the detection of hardcoded secrets in Android applications. In their work, the authors systematically categorized existing secret-detection approaches applicable to Android apps, covering more than ten existing tools~\cite{meli2019bad, gitleaks, repo-supervisor, trufflehog, whispers, ggshield, git-secrets-hunter, zuo2019does, feng2022automated, wen2022secrethunter, han2023credential, saha2020secrets, lounici2021optimizing}.  
They grouped these approaches into three main families of techniques:  
\wcircle{1} \emph{intrinsic-value-based analysis}, which relies on characteristics such as entropy and regular expression matching;  
\wcircle{2} \emph{static analysis}, which tracks data flows through APIs; and  
\wcircle{3} \emph{machine-learning-based methods}, which classify candidate strings using contextual features.  
Li et al. evaluated one representative tool from each family—Meli et al.~\cite{meli2019bad} for intrinsic-value analysis, \texttt{LeakScope}~\cite{zuo2019does} for static analysis, and \texttt{PassFinder}~\cite{feng2022automated} for machine-learning-based detection—on a dataset of \num{5135} Android apps, identifying \num{2142} secrets across \num{2115} distinct apps~\cite{li2024automaticallydetectingcheckedinsecrets}.  

Among these tools, only \texttt{LeakScope}~\cite{zuo2019does} was specifically designed for Android, while the others were adapted by Li et al. for mobile environments. However, \texttt{LeakScope} was limited to scanning for Google, AWS, and Microsoft cloud patterns and required significant computational resources.  
In a follow-up study, the same authors~\cite{wei2025far} conducted a large-scale empirical analysis using a regex-based approach, extracting hundreds of secrets (e.g., API keys and tokens) directly from released APKs, further demonstrating the prevalence of hardcoded credentials in the Android ecosystem.

\noindent
\textbf{Secrets Detection in Code.}
As also noted by Li et al.~\cite{li2024automaticallydetectingcheckedinsecrets}, most secret-detection tools were originally designed for generic source code rather than mobile applications. Extensive work has focused on identifying secrets in open-source repositories, particularly on platforms such as GitHub. Tools such as \textit{Gitleaks}~\cite{gitleaks} and \textit{TruffleHog}~\cite{trufflehog} scan Git histories using regular expressions and entropy heuristics to identify credentials such as AWS and Google API keys.  Commercial and enterprise-oriented tools, including \textit{GitGuardian’s ggshield}~\cite{ggshield} and \textit{SpectralOps}~\cite{spectralOps}, extend these techniques through large-scale pattern libraries and ML-assisted detection.  Basak et al.~\cite{basak2023comparative} evaluated several regex-based tools on open-source code and showed that such approaches generally suffer from low precision and recall. Academic research complements these practical efforts.  Meli et al.~\cite{meli2019bad} proposed a regex-based framework to analyze GitHub repositories for leaked cryptographic and API keys, while Feng et al.’s \textit{PassFinder}~\cite{feng2022automated} applied deep neural networks to detect password leaks by leveraging contextual embeddings rather than relying solely on string structures.  Similarly, Wen et al.~\cite{wen2022secrethunter} implemented a reinforcement-learning-based detection model, and El Yadmani et al.~\cite{el2025file} studied secret leakage in cloud APIs.  

Despite these advancements, all existing approaches share a common limitation: they depend on prior knowledge of what to detect, such as predefined regex patterns, known API structures, or labeled datasets. In contrast, \textit{SecretLoc} requires no such prior knowledge, leveraging large language models to infer secrets directly from contextual and structural cues.

\noindent
\textbf{Android App Security Analysis in General.}
Android security research broadly encompasses \textit{static}, \textit{dynamic}, and \textit{hybrid} analysis techniques. Among static approaches, \textit{FlowDroid}~\cite{arzt2014flowdroid} is one of the most widely adopted tools, capable of performing precise, context- and lifecycle-aware taint analysis for data leak detection in Android apps. Over the years, numerous other tools have been developed, including \textit{LeakDetector}~\cite{zhou2022uncovering}, \textit{ClipboardScope}~\cite{chen2024attention}, \textit{Amandroid}~\cite{wei2018amandroid}, and \textit{IccTA}~\cite{li2015iccta}, as well as several complementary frameworks and utilities~\cite{alecci2024improving, samhi2022difuzer, dex2jar, androguard, MobSF, apktool}. Dynamic analysis frameworks such as \textit{TaintDroid}~\cite{enck2014taintdroid} enable real-time instrumentation and runtime taint tracking, while other systems extend this paradigm to support monitoring and dynamic code analysis~\cite{zhauniarovich2015stadyna, kumar2023inviseal, lyons2023log}. Hybrid approaches combine both static and dynamic analysis to achieve higher coverage and accuracy, as exemplified by \textit{FSAFlow}~\cite{yang2022fsaflow}.
More recently, with the emergence of large language models (LLMs), researchers have begun integrating LLM-based reasoning into Android security analysis.
For example, \textit{AppPoet}~\cite{zhao2025apppoet} introduces a multi-view malware detection framework that combines static feature extraction with LLM-generated function descriptions and behavioral summaries, while \textit{LAMD}~\cite{qian2025lamd} leverages LLMs to extract security-critical contexts and construct structural representations for malware detection.
While these efforts collectively demonstrate significant progress in Android app security analysis, they primarily target vulnerabilities such as malware detection, user data leakage, or privacy violations. In contrast, our work focuses on a largely overlooked yet critical dimension—\textit{hardcoded secrets}—which introduces a distinct class of security risks within mobile apps.
\section{Conclusion}
\label{sec:conclusion}

In this paper, we presented \textsc{SecretLoc}, an LLM-based approach for detecting hardcoded secrets in Android apps. Unlike regex-, static-, or ML-based methods, \textsc{SecretLoc} leverages contextual reasoning to identify both known and previously unseen secrets without predefined patterns.
When evaluated on the current state-of-the-art benchmark for Android secret detection, \textsc{SecretLoc} successfully identified the vast majority of known secrets and additionally uncovered many new ones, including previously unreported categories such as OpenAI and Razorpay keys. These results demonstrate the ability of LLMs to generalize beyond fixed secret formats and detect obfuscated or unconventional credentials that evade traditional detection methods. When applied to \num{5000} recent Google Play apps, it revealed that over 40\% contained at least one exposed credential, confirming that secret exposure remains prevalent in production software.

Our results highlight the dual-use nature of LLM-based analysis: while \textsc{SecretLoc} can strengthen defensive auditing and foster better secret management, the same capability could be exploited for large-scale secret extraction. Responsible use and disclosure are therefore critical. Overall, our findings demonstrate that LLMs can significantly enhance secret detection, but also emphasize the need for secure-by-design development practices to prevent secret leakage.

\section{Ethics Considerations}
This study was conducted in accordance with responsible disclosure and ethical research principles. For the disclosure process, we contacted app developers using the official email addresses listed on Google Play, without including the actual secret values in the communication. 
Our goal was solely to notify developers of potential hardcoded secrets and to assist them in remediation if needed. As of this writing, several developers have confirmed that they are addressing or have already fixed the reported issues, and we continue to monitor responses to provide guidance when appropriate. No private or user data were collected or analyzed in this study.

\bibliographystyle{IEEEtran}
\bibliography{references.bib}

\end{document}